\documentclass[twocolumn,preprintnumbers,amsmath,amssymb,superscriptaddress]{revtex4}

\usepackage{epsf}
\usepackage{graphicx}
\usepackage{sidecap}

\usepackage{color}

\def \beq {\begin{equation}}
\def \eeq {\end{equation}}
\begin{document}

\title{{Distinct multiple fermionic states in a single topological metal}}


\author{M.~Mofazzel~Hosen}\affiliation {Department of Physics, University of Central Florida, Orlando, Florida 32816, USA}

\author{Klauss~Dimitri}\affiliation {Department of Physics, University of Central Florida, Orlando, Florida 32816, USA}

\author{Ashis~K.~Nandy}
\affiliation {Department of Physics and Astronomy, Uppsala University, P. O. Box 516, S-75120 Uppsala, Sweden}

\author{Alex~Aperis}
\affiliation {Department of Physics and Astronomy, Uppsala University, P. O. Box 516, S-75120 Uppsala, Sweden}

\author{Raman~Sankar} \affiliation{Center for Condensed Matter Sciences, National Taiwan University, Taipei 10617, Taiwan}
\affiliation{Institute of Physics, Academia Sinica, Taipei 10617, Taiwan}

\author{Gyanendra~Dhakal}
\affiliation {Department of Physics, University of Central Florida, Orlando, Florida 32816, USA}

\author{Pablo Maldonado}
\affiliation {Department of Physics and Astronomy, Uppsala University, P. O. Box 516, S-75120 Uppsala, Sweden}

\author{Firoza Kabir}
\affiliation {Department of Physics, University of Central Florida, Orlando, Florida 32816, USA}

\author{Christopher Sims}
\affiliation {Department of Physics, University of Central Florida, Orlando, Florida 32816, USA}

\author{Fangcheng Chou} \affiliation{Center for Condensed Matter Sciences, National Taiwan University, Taipei 10617, Taiwan}

\author{Dariusz Kaczorowski}
\affiliation {Institute of Low Temperature and Structure Research, Polish Academy of Sciences,
50-950 Wroclaw, Poland}

\author{Tomasz~Durakiewicz}
\affiliation {Condensed Matter and Magnet Science Group, Los Alamos National Laboratory, Los Alamos, NM 87545, USA}

\author{Peter M.\ Oppeneer}
\affiliation {Department of Physics and Astronomy, Uppsala University, P. O. Box 516, S-75120 Uppsala, Sweden}

\author{Madhab~Neupane}
\affiliation {Department of Physics, University of Central Florida, Orlando, Florida 32816, USA}

\date{\today}
\begin{abstract}
%

Among the quantum materials that gained interest recently are the topological Dirac/Weyl semimetals, where the conduction and valence bands touch at points in reciprocal (\textit{k})-space,  and the Dirac nodal-line semimetals, where these bands touch along a line in $k$-space.   
However, the coexistence of multiple fermion phases in a single material has not been verified yet. 
Using angle-resolved photoemission spectroscopy (ARPES) and first-principles electronic structure calculations, we systematically study the 
metallic topological quantum material, Hf$_2$Te$_2$P. Our investigations discover various properties which are rare and never observed in a single Dirac material.
We observe the coexistence of both weak and strong topological surface states in the same material and interestingly, at the same momentum position.
A one-dimensional Dirac crossing--the \textit{Dirac-node arc}--along a high-symmetry direction is revealed by our first-principles calculations and confirmed by our ARPES measurements. 
This novel state is associated with the surface bands of a weak topological insulator protected by \textit{in-plane time-reversal invariance}.
Ternary compound Hf$_2$Te$_2$P thus emerges as an intriguing platform to study the coexistence and competition of multi-fermionic states in one material.

\end{abstract}
\maketitle


The discovery of topological insulators (TI) \cite{Hasan, SCZhang, Xia, Hasan_review_2, Neupane_4} has invigorated intense research efforts in the quest for novel non-trivial surface states (SSs) in a wider group of quantum materials \cite{Dai, Neupane_2,Nagaosa, Young_Kane,Suyang_Science}, which includes bulk insulators, semimetals, and metals. 
These appearing surface states can manifest themselves from multiple origins such as strong spin-orbit coupling (SOC), accidental band touching, symmetry protection, strong electronic correlation effects, etc \cite{Neupane_4, BiSe, BiSe1, Neupane, Ti2}. 
Uncovering the origin of topological protection of the surface states has hence become  an important research topic.

A tetradymite-type Bi-based component was the first experimentally discovered TI in which the surface state was realized through the SOC effect and protected by time-reversal symmetry (TRS) \cite{Hasan, SCZhang}. 
Moving from topological surface states (TSSs) in the TIs, topological Dirac semimetals (TDSs) such as Cd$_3$As$_2$ \cite{Dai, Neupane_2, bori} and Na$_3$Bi \cite{Na1, Na2, Na3, Na4} were discovered to have band touchings between the valence and conduction bands at certain discrete $k$-points in the Brillouin zone (BZ) and naturally host linear Dirac dispersions in three-dimensional (3D) momentum space. 
These TDSs with four-fold degeneracy are protected by additional crystalline symmetries other than the TRS and inversion symmetry (IS) \cite{Dai, Na1, Na2, 3D-DS}.
Another subsequently discovered semimetallic class, that of Weyl semimetals, is featured with pairs of bulk Dirac-cones and the connecting Fermi-arc SS \cite{TaAs_theory,TaAs_theory_1,Suyang_Science, Hong_Ding}. 
The metallic surface states of the Weyl semimetals show photon-like linear band dispersion and exhibit a variety of exotic properties that include large magnetoresistance and high carrier mobility \cite{TaAs_theory,TaAs_theory_1,Suyang_Science, Hong_Ding}. 
Evolving from the Weyl semimetals, nodal-line semimetals (an extended TDS class) possess a 1D bulk band touching in a loop in momentum space and are accompanied by so-called `drumhead' surface bands \cite{node_0, PTS, Schoop, Neupane_5}.
Here, one needs additional crystalline symmetries to protect the degeneracy of the line-node against perturbation, such as mirror symmetry \cite{PTS}, or nonsymmorphic symmetry \cite{ Schoop, Neupane_5, ZrSiX}. 
The high density of states in the topological 1D fermion state provide a possible route to discover materials that could exhibit high temperature surface superconductivity \cite{Kopnin_PRB}.

\begin{figure*}
	\centering
	\includegraphics[width=17.5cm]{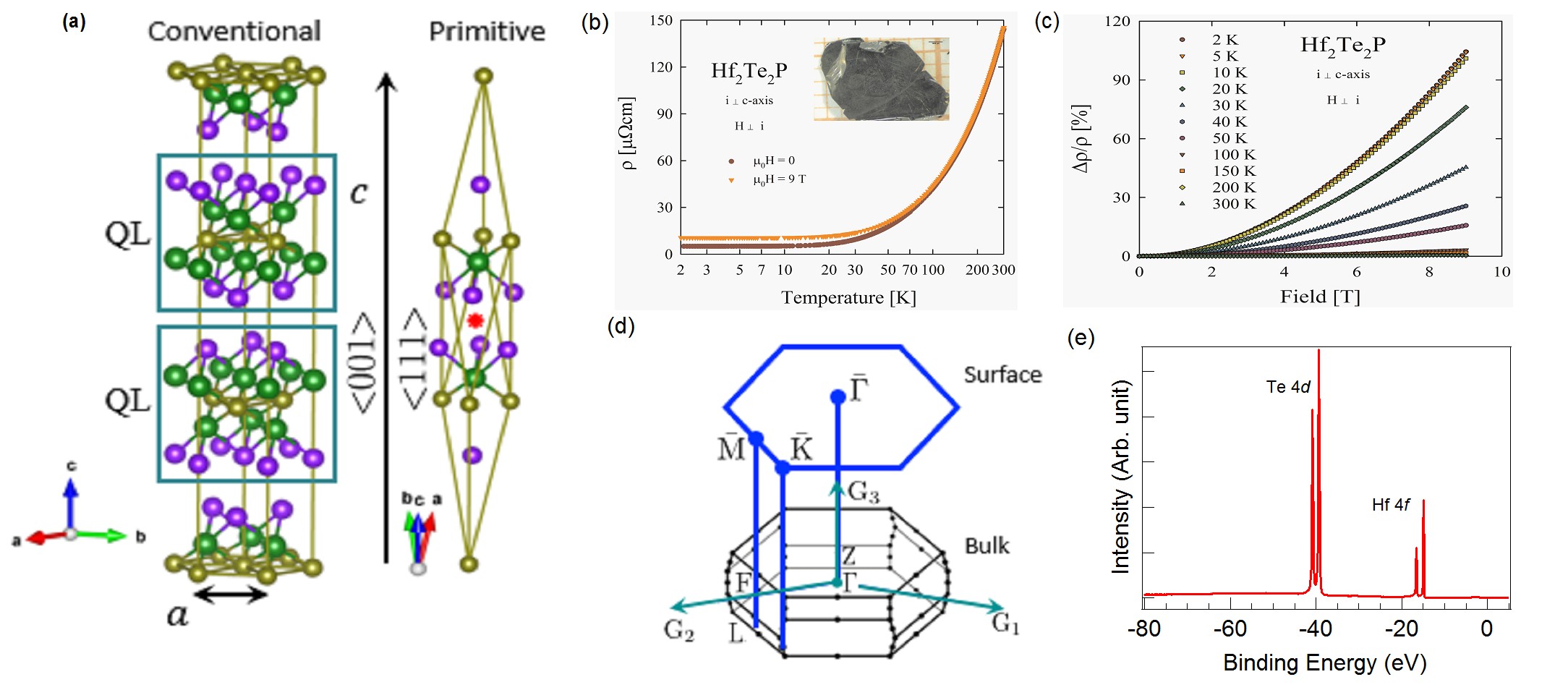}
	\caption{\textbf{{Crystal structure and sample characterization of Hf$_2$Te$_2$P}.} 
	\textbf{a,} The rhombohedral tetradymite crystal structure of Hf$_2$Te$_2$P, depicted in the conventional (hexagonal) and primitive (rhombohedral) unit cells.   The stacking of quintuple layers (QL) is depicted by the blue square. The inversion center is indicated with the red star. 
\textbf{b,} Temperature dependence of the electrical resistivity measured on a single crystal of Hf$_2$Te$_2$P in a zero and 9 T magnetic field applied perpendicular to the current flowing within the basal plane of the crystallographic unit cell. The inset shows a picture of one of the single crystal grown for the present research.
\textbf{c,} Magnetic field dependencies of the transverse magnetoresistance  measured for a range of different temperatures on a single crystal of Hf$_2$Te$_2$P. 
\textbf{d,} 3D bulk Brillouin zone and its projection on the hexagonal surface Brillouin zone of the Hf$_2$Te$_2$P-crystal. High symmetry points are marked in the plot.
		\textbf{e,} Core level spectroscopic measurement of Hf$_2$Te$_2$P. Sharp peaks of Te 4\textit{d} and Hf 4\textit{f} are observed which indicate good sample quality.}
\end{figure*}

In spite of the already discovered materials hosting one of the mentioned quantum phases, materials that exhibit several TSSs at distinct reciprocal space locations would be highly desirable as these could unlock an interplay of topological quantum phenomena. 
Here, through the use of ARPES and \textit{ab initio} calculations, we report the discovery of topologically distinct fermionic SSs, multiple Dirac crossings, and the \textit{Dirac-node arc}, in a single topological metal Hf$_2$Te$_2$P, a compound from the 221 family.
This family of materials has recently drawn interest; replacing Hf by a lighter atom Zr, one obtains Zr$_2$Te$_2$P which is a strong topological metal with multiple Dirac cones \cite{ZTP_PRB}.
Unlike the closed loop line-nodes accompanied with `drumhead' surface states, here, the surface band crossing, resembling the Dirac state in graphene, is extended in a short line along a  high-symmetry direction.  
Our ARPES measurements reveal the presence of multiple pockets at the Fermi level (\textit{E$_\textrm{F}$}) collectively comprising of a 6-fold flower with ``petal"-shaped Fermi surface (FS). 
Apart from several Dirac cones located at the $\Gamma$ point at different binding energies (\textit{E$_\textrm{B}$}), below and above the Fermi level, our most important observation is that of the \textit{Dirac-node arc} seen here
in a topological metal, centered at the M point along the $\Gamma$-M-$\Gamma$ direction.
The experimental observations are explained by our band structure calculations within density functional theory (DFT). 
In contrast to the \textit{Dirac-node arc} of unambiguous origin observed in topological line-node semimetals \cite{HSS}, our DFT study reveals that here it is a signature of a weak topological \textit{Z$_2$} invariants and protected by \textit{in-plane time-reversal invariance} \cite{Lau_PRB}. Consequently, this material could be the first system to realize the coexistence of both weak and strong \textit{Z$_2$} invariants due to the presence of multiple bulk topological bands.


\begin{figure*}
	\centering
	\includegraphics[width=17.5cm]{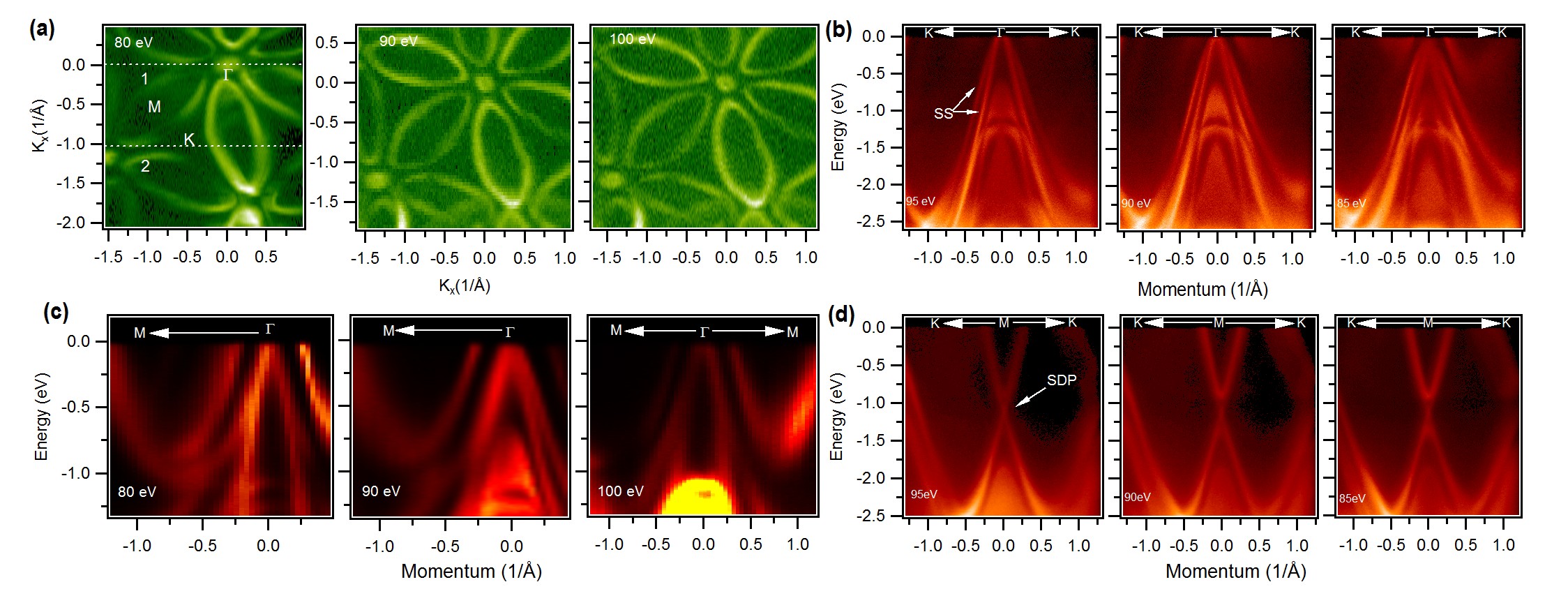}
	\caption{\textbf{Fermi surface and observation of multiple fermionic states.} 
		\textbf{a,} Fermi surface maps at various photon energies. Photon energies are marked on the plots. The white dashed lines marking No.\ 1 and 2 denote the direction of the dispersion maps. \textbf{b-d,} Dispersion maps measured along various high-symmetry directions for different photon energies. These data were collected at the SIS-HRPES end station at the SLS, PSI at a temperature of 18 K.}
\end{figure*}
\bigskip

\textbf{Crystal structure and sample characterization}\\
The materials of the 221 family are of particular interest due to their tetradymite-type layered crystal structure, mostly found in typical 3D topological insulators as e.g., Bi$_2$Te$_3$, Bi$_2$Se$_3$, and Sb$_2$Te$_3$ \cite{Neupane_4, BTS1, ZTP_PRB, ZPT, HT}, having a three-fold rotation symmetry about the $z$-axis. 
Similar to the well known topological insulator Bi$_2$Te$_2$Se \cite{BTS2}, Hf$_2$Te$_2$P crystallizes in a rhombohedral crystal structure with space group $R\bar{3}m$ (No.\ 166) \cite{crystal}. 
The conventional (hexagonal) unit cell consists of three basic quintuple layers, each with stacking sequence Te-Hf-P-Hf-Te (see Fig.\ 1(a)).
The atoms within a quintuple layer are covalently bonded, whereas between layers they are bonded by weak van der Waals forces which facilitate to cleave along the $\{001\}$ basal plane, as shown in Fig.\ 1(a).
The primitive (rhombohedral) unit cell possesses an inversion symmetry depicted by the red star.  
The crystal also respects the TRS as well as preserve the reflection symmetry with respect to the mirror planes $\{110\}$ and $\{100\}$.
This reflection symmetry may play a distinct role in conserving \textit{in-plane time-reversal invariance} and hence, protecting the \textit{Dirac-node arc}, as discussed later. 
Fig.\ 1(b) shows the temperature dependence of the electrical resistivity of Hf$_2$Te$_2$P which indicates the metallic nature of this material.
The transverse magnetoresistance MR = [$\rho$(T,H) - $\rho$(T,0)]/$\rho$(T,0) attains a value of about 100\% below 10 K. 
At higher temperatures, the magnitude of MR rapidly decreases and drops below 1\% above 150 K. 
The bulk and (111)-projected surface BZs of the conventional unit cell are shown in Fig.\ 1(d). 
The high symmetry points $\Gamma$, Z, F, and L are on the mirror planes of either \{110\} or \{100\} type in the bulk BZ.
The projected surface BZ is perpendicular to the mirror planes.
The center of the surface BZ is denoted as the $\Gamma$ point, the K points are at the corners of the surface BZ, and M is the mid-point of two adjacent corners. 
Fig.\ 1(e) shows the spectroscopic core level measurements of Hf$_2$Te$_2$P. 
We observe sharp peaks of Hf 4\textit{f} at around 14.2 eV and 15.9 eV, and Te 4\textit{d} at around 40 eV. 
This indicates that the sample used in our measurements is of good quality. \\

\textbf{Fermi surface, Dirac nodes, and \textit{Dirac-node arc}}\\
We now report the experimental results that reveal the details of the electronic structure including the Fermi surface and the TSSs with multiple Dirac-like features. 
To this end we probed the electronic structure of Hf$_2$Te$_2$P through the use of ARPES with low incident photon energy (80--100 eV). 
Fig.\ 2(a) shows the Fermi surface maps with ARPES intensity integrated in a 20-meV energy window for Hf$_2$Te$_2$P at various photon energies. 
It reveals the presence of multiple Fermi surface pockets such as the ``petal"-shaped 6 electron-like pockets along the $\Gamma$-M directions and a hole-like pocket at the zone center $\Gamma$, resembling a 6-fold flower-like Fermi surface. 
This type of pronounced electronic dispersion is extremely rare in Dirac-type materials. 
The 6-fold flower petal-like Fermi surface reflects the 3-fold rotational  and inversion symmetry of the lattice. 

Fig.\ 2(b)-(d) show the energy dispersion maps along various key directions indicated on the Fermi surface of Fig.\ 2(a) (see also Supplementary Information).
Fig.\ 2(b) and (c) show the dispersion maps along the K-$\Gamma$-K and M-$\Gamma$-M directions, respectively, at different photon energies as denoted on the plots.
A clear linearly dispersive state (Dirac-like) is observed around the $\Gamma$ point with a wider range of band dispersion.
The chemical potential of the Dirac crossing is found at the top of this highly linearly dispersive state.
In Fig.\ 2(c), the linearly dispersive states along the $\Gamma$-M direction do not disperse with photon energy. 
The linear dispersion may have both bulk and surface features at the zone center of the BZ. 
Furthermore, guided by the theoretical calculations, we carefully analyze the dispersion map along the K-$\Gamma$-K and at \textit{E$_\textrm{B} \approx$} 1.2 eV, where we observe a second Dirac cone-like feature.
However, the Dirac cone is not clearly observed due to the broad ARPES intensity, therefore, we confirmed it by taking a 2D curvature plot of this dispersion map (see Supplementary Information).
Rather, in Fig.\ 2(d), an extremely sharp Dirac-like linear dispersion within a wide energy window is observed along the K-M-K direction for all three photon energies, and therefore more likely to be due to the SSs.
Importantly, the linearly dispersive energy range is as large as 2.2 eV below and above the Fermi level which is even larger than the recently reported nodal-line semimetal ZrSiS \cite{Schoop, Neupane_5}.
Finally, the Dirac-like band touching is observed at the M point at \textit{E$_\textrm{B} \approx$} 1.1 eV.
Importantly, our ARPES results establish the presence of multiple Dirac surface features (two crossings) located at different binding energies at different high-symmetry points $\Gamma$ and M.
The even number of Dirac features within a bulk pseudogap is a clear signature of weak \textit{Z$_2$} invariants in the Hf$_2$Te$_2$P material.
The Dirac-like feature close to the M point is much broader than the Dirac-cone around the zone center $\Gamma$. 
\begin{figure*}
	\centering
	\includegraphics[width=17.5cm]{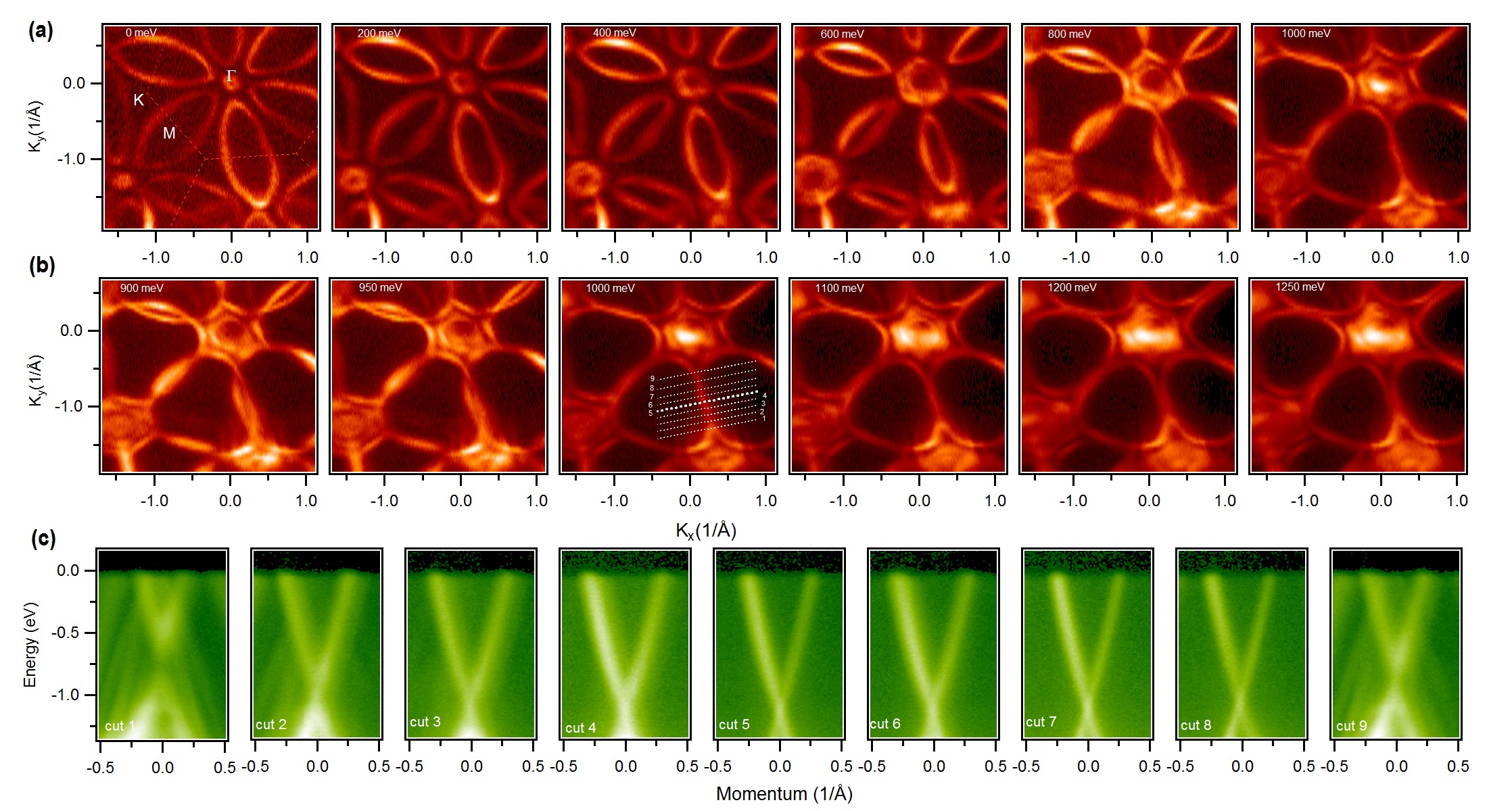}
	\caption{ \textbf{Experimental observation of the \textit{Dirac-node arc}.}
		\textbf{a,} Constant energy contour plots at various binding energies. \textbf{b,} Constant energy contour plots closer to the \textit{Dirac-node arc}. Binding energies are given in the plots. \textbf{c,} Dispersion map along the K-M-K direction along the cut directions indicated in the 1000-meV constant energy contour panel of Fig.\ 3b. All data were collected at the SIS-HRPES end station at the SLS, PSI at a photon energy of 100 eV with a temperature of 18 K.}
\end{figure*}

Now we report our most important ARPES observation, of the \textit{Dirac-node arc}, as a signature of weak TI.
Fig.\ 3(a) shows the constant energy contours at a photon energy of 100 eV as a function of binding energy \textit{E$_\textrm{B}$}, as noted on the plots.  
Moving towards higher \textit{E$_\textrm{B}$}, we observe that the ellipsoidal ``petal"-shaped features are gradually shrinking in width along the K-M-K direction whereas the small circular pocket at the $\Gamma$ point is increasing in size. 
This indicates the hole-like nature of the pocket at the zone center and the electron-like nature of the ellipsoidal pockets. 
At \textit{$E_\textrm{B} \approx $} 1 eV, we observe that the ellipsoidal-like pocket disappears and finally forms a 1D line-like feature along the high-symmetry M-$\Gamma$-M direction.
A closer look in energy, Fig.\ 3(b), shows that the novel line-like fermionic state begins to flatten at a binding energy of 1.1 eV.   
Next, we focus on the energy dispersion around the M point in more detail according to the cuts shown in the momentum dispersion curve at \textit{E$_\textrm{B}$} = 1 eV, Fig.\ 3(b).
Along cuts No.\ 1 to No.\ 9, the band dispersions shown in Fig.\ 3(c) describe the detailed evolution of the Dirac features along the K-M-K direction.
At the M point in cut No. 5,\ a very sharp linear dispersion starts to cross at \textit{E$_\textrm{B}$} = 1 eV, see also Fig.\ 2(d). 
The Dirac crossing moves up in energy on both sides of the M point, making cuts No.\ 2-4 and cuts No.\ 6-8 symmetrical. Importantly, each crossing resembles the Dirac state in graphene.
Therefore, in the proximity of TRS point M the movement of such Dirac crossing forms a line, the \textit{Dirac-node arc} in energy-momentum space. 
Moving further away from the M point, cuts No.\ 1 and 9 show a slightly gapped Dirac cone with a largely linear dispersion and the bulk states dominate over the surface contribution.

\begin{figure*}
  	\centering
 	\includegraphics[width=16.5cm]{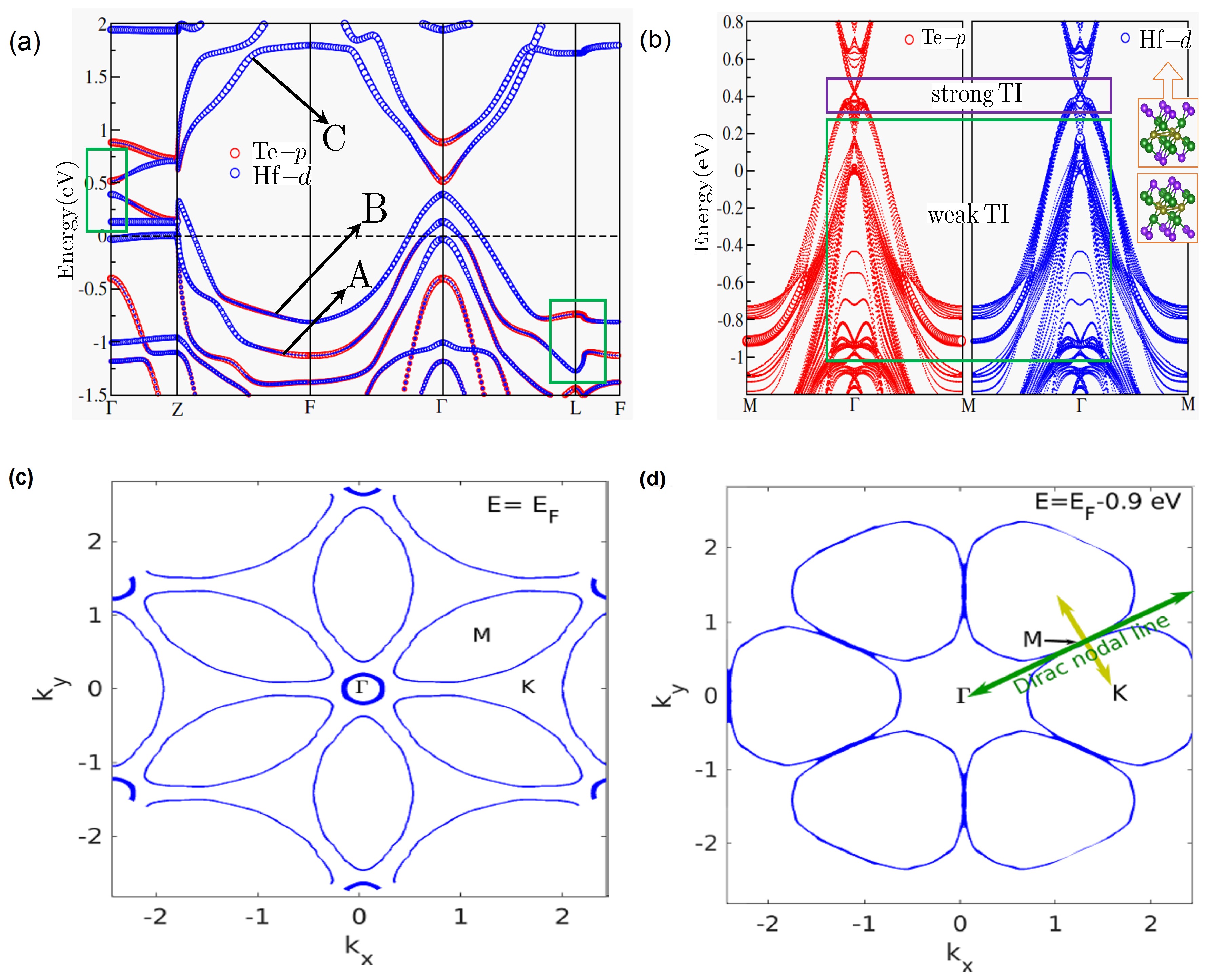}
  	\caption{\textbf{Calculated multiple fermionic states.}
  		\textbf{a,} The bulk electronic structure, calculated with spin-orbit coupling, along high-symmetry directions. Topologically protected Dirac cones are marked with the green rectangles. Blue and red circles indicate the Hf-\textit{d} and Te-\textit{p} character of the band states, respectively. \textbf{b,} Calculated Hf and Te terminated surface state for the (111) surface. \textbf{c,\,d,} Calculated Fermi surface and constant energy contour 900 meV below the Fermi level, respectively. At 900 meV below the Fermi level we observe a symmetry protected Dirac nodal-line state, denoted by the green arrow.}
 \end{figure*}
\bigskip

\textbf{Bulk and surface states}\\
For a detailed understanding of the novel ARPES features, we now focus on the electronic structure calculations of Hf$_2$Te$_2$P based on DFT. 
All the observed features are well explained by our \textit{ab initio} calculations which guided us to identify the Dirac-node arc, and disentangle bulk and surface origins. 
Fig.\ 4(a) shows the bulk electronic structure of Hf$_2$Te$_2$P along the high-symmetry directions including four time-reversal-invariant momenta (TRIM), calculated with spin-orbit coupling.
Several bands cross the Fermi energy that are formed of hybridized Hf-$d$ (blue circles) and Te-$p$ (red circles) states; the size of the circles indicates the contribution for a specific orbital.
The bands close to \textit{E$_\textrm{F}$} around the F point are almost ``flat bands", i.e., they are approximately dispersionless and for the sake of simplicity, these are labeled as bands A, B, and C which are predominantly formed of Te-$p$, Te-$p$ and Hf-$d$ orbital characters, respectively.
Irrespective of the \textit{E$_\textrm{F}$} position, direct energy gaps exist between the band sets (A,B) and (B,C) in the whole BZ, owing to the SOC; this direct energy gap determines the probable topology of SSs.
Apart from a clear band inversion of the $d\!-\!p$ type between bands B and C at the $\Gamma$ point, there is an additional inversion among the topmost valence bands at the L point induced by SOC, highlighted by the green ellipses.
These band inversions are usually considered as origin of the topological nontriviality i.e., nontrivial TSSs.
As shown in Fig.\ 1(a), the primitive unit cell with the inversion center at the P atom further allows us to determine the parity of the Bloch wave function which is consistent with that of the corresponding atomic orbital: `+' for the $s$ and $d$ orbitals and `$-$' for the $p$ orbital.
The $d$ and $p$ parity orders at the Z and L points are trivial between bands A and B and hence, a trivial $Z_2$ topological invariant is expected.
However, considering band set (B,C) it is clear that Te-$p$ and Hf-$d$ bands become inverted at the $\Gamma$ point with respect to the orders in other TRIM points, leading to a nontrivial $Z_2$ number.    
To confirm the topological origin of the observed features in the ARPES experiment, we calculate the four topological $Z_2$ invariants, ($\nu_0\!:\!\nu_1,\!\nu_2,\!\nu_3$), assuming \textit{E$_\textrm{F}$} can be tuned into the pseudogap between band sets (A,B) and (B,C).
This calculation is based on considering the parity of the predominant orbital character at the four irreducible TRIM points of the BZ, as described by Fu and Kane \cite{Fu&Kane}.    
Interestingly, within the band set (A,B), we find $\nu_0$ = 0 but the other topological invariants are calculated to be nonzero, ($\nu_1,\!\nu_2,\!\nu_3$) = (1,1,1), $-$ a weak topological $Z_2$ invariant.
Once we consider the pseudogap between bands B and C, strikingly, $\nu_0$ changes from 0 to 1 due to the band inversion at $\Gamma$, while weak invariants ($\nu_1,\!\nu_2,\!\nu_3$) remain unaltered $-$ a strong topological $Z_2$ invariant.  
Therefore, Hf$_2$Te$_2$P is a unique quantum material carrying multiple topological bands along with both weak and strong $Z_2$ invariants.
Both $Z_2$ topological invariants guarantee the existence of TSSs on the (111) surface.

Indeed, in our Te terminated (111) surface calculations, we observe multiple topologically protected Dirac states, see Fig.\ 4(b). 
First of all, a pair of linearly dispersive TSSs exist around the $\Gamma$ point inside the $d-p$ inversion gap (between bulk band set (B,C)) with the Dirac point lying at about 0.5 eV above the \textit{E$_\textrm{F}$}. 
At about 0.17 eV above \textit{E$_\textrm{F}$}, another Dirac point is observed at the same $\Gamma$ point protected by the band topology of bulk states A and B. 
Here, the surface character of the Dirac dispersion is strongly mixed with the bulk character. 
Unfortunately, the above mentioned two Dirac points cannot be visualized in our ARPES measurements as only occupied electronic bands can be probed by this technique.
However, our constant energy contour plots in Fig.\ 4(c),(d) confirm that the first Dirac point is derived from the sharp corner of the 6 ``petal"-shaped pockets meeting at a point above \textit{E$_\textrm{F}$} while the later one is derived from the circular ring at the zone center as observed in ARPES, see Fig.\ 2(a).   
In stark contrast to a single Dirac point at 0.5 eV above \textit{E$_\textrm{F}$}, a nontrivial line-like feature is observed around \textit{E$_\textrm{B}$} $\sim$ 0.9 eV at the M point along the $\Gamma$-M-$\Gamma$ direction originating from the same band topology of the bulk band set (A,B).
Such highly anisotropic bands around the M point provide the possibility of additional tunability in this material \cite{AMnBi2, Sr_CaMnBi2} as electrons can propagate differently from one direction to another.
The calculated bands along the K-M-K direction show excellent agreement with our experimentally measured dispersion maps (see Fig.\ 3(a)). 
We find that the energy window of linear dispersion is more than 2.3 eV, extending both below and above the Fermi level, which is more than any known topological material. 
Our calculations reveal an even number of topological nontrivial states, a Dirac cone at $\Gamma$ and one at M, and the \textit{Dirac-node arc} along the M-$\Gamma$ direction, all due to the bulk band set (A,B), $-$ a clear signature of a weak $Z_2$ invariant. 
Furthermore, besides the strong and weak TI states above the chemical potential, we also observe a Dirac-like state at the same momentum position, $\Gamma$, at about \textit{E$_\textrm{B}$} = 1.1 eV, consistent with the ARPES observation (see Supplementary Information).
However, the topological origin of this state is unclear and possibly it is a bulk Dirac feature. 


 \textbf{Discussion}\\
First of all, we observe a 6-fold flower ``petal"-shaped Fermi surface which shows that even in a metallic Dirac material such a remarkable dispersion is possible. 
Secondly, we observe multiple Dirac cones with linear dispersion over a wide energy range ($\sim$2.3 eV), even larger than that of ZrSiS ($\sim$2 eV). 
Most importantly, in the well-studied typical $n-$type Bi$_2$Se$_3$/Bi$_2$Te$_3$ \cite{Hasan, SCZhang} topological insulator materials (Fig. 5(a)) it is experimentally observed that the Dirac cones have lower and upper cone giving the Fermi level well above the Dirac point. On the other hand, for the distinct $p-$type material such as Sb$_2$Te$_3$ \cite{Hasan, SCZhang}, the Dirac point is located well above the Fermi level (Fig. 5(b)).  Furthermore, in ZrSiS-type nodal line materials, the conduction and valence band touch each other along a one dimensional loop or line protected by nonsymmorphic symmetry \cite{Schoop,Neupane_5} (Fig. 5(c)).
Thus the three phenomena described are uniquely found in a distinct material or family of materials. Never before has a single material hosted three such topological states. Incredibly, we have acquired sufficient experimental and theoretical evidence to report that Hf$_2$Te$_2$P hosts multiple Dirac states (Fig. 5(d-e)).

We speculate that there might be competition between these two states as we observe that the \textit{Dirac-node arc} appearing around the M point is squeezed to very small reciprocal space. 
The Dirac features (arc and node at the M and $\Gamma$ point, respectively) originating from the topological bulk band set (A,B) are protected by the in-plane time-reversal symmetry $-$ a 2D analog of the conventional time-reversal symmetry \cite{Lau_PRB}.
The \{110\} and/or \{100\} mirror planes in the Hf$_2$Te$_2$P crystal confirm the presence of in-plane time-reversal symmetry and according to the Ref.\ \cite{Lau_PRB}, one finds topologically protected Dirac lines on the surface of weak topological invariants, here it is the (111) surface. 
Interestingly, we observe that one of the Dirac nodes at the $\Gamma$ point is weak and the other one is a strong topological state with $Z_2$ invariants (0:111) and (1:111), respectively. 
In contrast to the other known topological materials our first-principles calculations show that the band inversion is \textit{d-p} type similar with Zr$_2$Te$_2$P \cite{ZTP_PRB} instead of a \textit{s-p} type band inversion. 
The bands around the M points are highly anisotropic so there is a large possibility of a tunable Dirac cone state that may lead to new properties. 
This system provides an unique opportunity to study multiple fundamental fermionic quantum phases in the same material.    

 \begin{figure*}
  	\centering
 	\includegraphics[width=16.5cm]{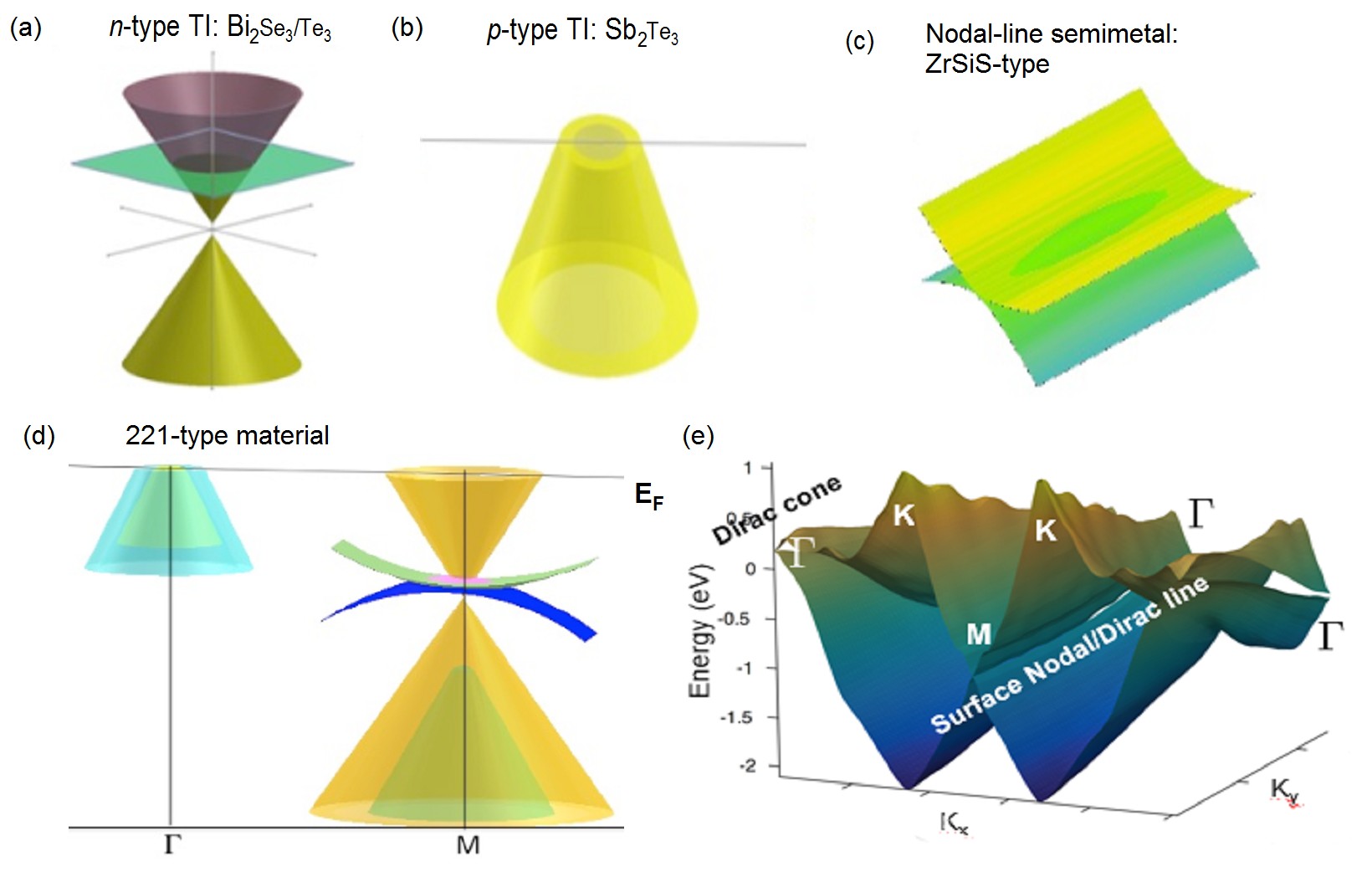}
 	\caption{\textbf{Schematic view of distinct fermionic states.}
 		\textbf{a-c,} Electronic dispersion of the $n$-type topological insulator, $p$-type topological insulator and nodal-line semimetal, respectively. \textbf{d,} Sketch of electronic dispersions of the 221-material Hf$_2$Te$_2$P. This material consists of both $n$- and $p$-type topological surface states as well as a nodal-line semimetal phase. \textbf{e,} Schematic view of the calculated electronic structure of this material. }
 \end{figure*}
\bigskip




 \bigskip
 \textit{Note added}: During the preparation of our manuscript, we noticed that Ref. \cite{221} also discussed topological property of 221 system.
 
 \bigskip

 \textbf{Methods}
 \bigskip
 
 \textbf{Crystal growth}
 
 The single crystals of Hf$_2$Te$_2$P were grown through the vapor transport method as described elsewhere \cite{ZPT}. The crystal structure was determined by X-ray diffraction on a Kuma-Diffraction KM4 four-circle diffractometer equipped with a CCD camera using Mo K$\alpha$ radiation, while chemical composition was checked by energy dispersive X-ray analysis performed using a FEI scanning electron microscope equipped with an EDAX Genesis XM4 spectrometer.
 Electrical resistivity measurements were carried out within the temperature range 2-300 K and in applied magnetic fields of up to 9 T using a conventional four-point ac technique implemented in a Quantum Design on a four-circle PPMS platform. The electrical contacts were made using silver epoxy paste. 
 \bigskip
 
 \textbf{Spectroscopic characterization}
 
  Synchrotron-based ARPES measurements were performed at the SIS-HRPES end-station at the Swiss Light Source (SLS) and Advanced Light Source (ALS) beamline 4.0.3 with  Scienta R4000  and R8000 hemispherical electron analyzers. The angular resolution was set to be better than 0.2$^{\circ}$, and the energy resolution was set to be better than 20 meV for the measurements. The samples were cleaved \textit{in-situ} under vacuum condition better than 3$\times$10\textsuperscript{-11} torr and at a temperature around 16 K. 
 \bigskip
 
\textbf{Electronic structure calculations}

The electronic structure calculations and structural optimization were carried out within the density-functional formalism as implemented in the Vienna \textit{ab initio} simulation package (VASP) \cite{Kresse_1, Kresse_2}.
Exchange and correlation were treated within the generalized gradient approximation (GGA) using the parametrization of Perdew, Burke, and Ernzerhof (PBE) \cite{GGA_1}.
The projector-augmented wave (PAW) method \cite{PAW, Blochl} was employed for the wave functions and pseudopotentials to describe the interaction between the ion cores and valence electrons.
The lattice constants and atomic geometries were fully optimized and obtained by minimization of the total energy of the bulk system. 
The surface of the two-dimensional crystals was simulated as a slab calculation within the supercell approach with sufficiently thick vacuum layers.
In addition to the general scalar-relativistic corrections in the Hamiltonian, the spin-orbit interaction was taken into account.
The plane wave cutoff energy and the $k$-point sampling in the Brillouin zone integration were checked carefully to assure the numerical convergence of self-consistently determined quantities.

 \bigskip
\textbf{Acknowledgment}\\
 M.N.\ is supported by the start-up fund from the University of Central Florida.
 T.D.\ is supported by the NSF IR/D program. 
 D.K.\ was supported by the National Science Centre (Poland) under research grant 2015/18/A/ST3/00057.
 A.K.N., A.A., P.M., and P.M.O.\ acknowledge support from the Swedish Research Council (VR), the K. and A. Wallenberg Foundation (Grant No.\ 2015.0060) and the Swedish National Infrastructure for Computing (SNIC). We thank Plumb Nicholas Clark for beamline assistance at the SLS, PSI. We also thank Jonathan Denlinger for beamline assistance at the LBNL.

\end{document}